\documentclass[reprint, superscriptaddress, nofootinbib, pra, aps]{revtex4-2}
\usepackage{amsmath, amsfonts, amssymb, bbm, physics}
\usepackage{color, enumitem, graphicx, subcaption}
\usepackage[hidelinks]{hyperref}
\usepackage{amsthm}
\newtheorem{lemma}{Lemma}[section]

\usepackage{geometry}
\usepackage{array, float}
\newcolumntype{M}[1]{>{\centering\arraybackslash$}m{#1}<{$}}
\title{\textbf{{James-Stein Estimation in Guassian Metrology : V2}}}

\renewcommand{\th}[1][]{ \ensuremath{  \hat{\theta}^{\text{#1}}  } }
\def\MLE{\ensuremath{\th[MLE]}}
\def\JS{\ensuremath{\th[JS]}}
\def\mJS{\ensuremath{\th[mJS]}}
\def\tb{\ensuremath{\th[B]}}
\newcommand{\wt}[1]{\ensuremath{\widetilde{#1}}}

\def\Tr{\ensuremath{\text{Tr}}}
\def\AD{\ensuremath{\text{AD}}}
\def\PAD{\ensuremath{\text{PAD}}}

\usepackage{soul}

\begin{document}
\def\DAMTP{DAMTP, Centre for Mathematical Sciences, University of Cambridge, United Kingdom}
\def\HCL{Hitachi Cambridge Laboratory, J. J. Thomson Avenue, CB3 0HE, Cambridge, United Kingdom}
\def\Warwick{Department of Computer Science, University of Warwick, United Kingdom}

\author{Wilfred Salmon}
\affiliation{\DAMTP}
\affiliation{\HCL}
\author{Sergii Strelchuk}
\affiliation{\DAMTP}
\affiliation{\Warwick}
\author{David R. M. Arvidsson-Shukur}
\affiliation{\HCL}

\title{James-Stein Estimation in Quantum Gaussian Sensing}

\begin{abstract}
The James-Stein estimator is a biased estimator---for a finite number of samples its expected value is not the true mean. The maximum-likelihood estimator (MLE), is unbiased and asymptotically optimal. Yet, when estimating the mean of $3$ or more normally-distributed random variables, the James-Stein estimator has a smaller total (expected) error than the MLE. We introduce the James-Stein estimator to the field of quantum metrology, from both the frequentist and Bayesian perspectives. We characterise the effect of quantum phenomena on the James-Stein estimator through the lens of quantum Gaussian sensing, the task of estimating the mean of an unknown multivariate quantum Gaussian state. We find that noiseless entanglement or coherence improves performance of the James-Stein estimator, but diminishes its advantage over the MLE. In the presence of noise, the James-Stein advantage is restored. Quantum effects can also boost the James-Stein advantage. We demonstrate this by investigating multivariate postselective metrology (generalised weak-value amplification), a strategy that uses quantum effects to measure  parameters with imperfect detectors. Simply by post-processing measured data differently, our techniques reduce errors in quantum experiments.
\end{abstract}

\maketitle

\section{Introduction}
The James-Stein estimator underlies one of the most counterintuitive phenomena in classical statistics \cite{Stei1956, Stei1961}. Suppose one wishes to estimate 3 or more independent normally distributed quantities, minimising the total mean-squared error. The most common strategy for such a task is the maximum-likelihood estimator (MLE), which treats each quantity separately. However, Stein \cite{Stei1956} showed the surprising result that it is always better to combine the data for each quantity, \textit{although the underlying quantities are completely independent}. Such estimation strategies are now referred to as James-Stein estimators \cite{Stei1961}. We refer to the advantage of the James-Stein estimator over the MLE as the James-Stein advantage. \\

In this Article, we consider the application of James-Stein estimation to parameters encoded in quantum systems. We focus on the particular case of Gaussian shift models \cite{Demk2020}, which are the quantum generalisations of a Gaussian distribution with an unknown mean. The central-limit theorem \cite{Wass2004} ensures that the mean of many identical, independent measurements of any parameterised state is approximately Guassian. Moreover, quantum local asymptotic normality \cite{Demk2020} implies that many copies of any parameterised quantum state may be locally approximated as a Gaussian shift model. Therefore, the study of Gaussian shift models is ubiquitous in quantum metrology. We analyse the James-Stein advantage in  commonly studied areas of quantum metrology: noiseless metrology, noisy metrology, Bayesian metrology and metrology with postselection (all defined below). We find that boosting metrology by noiseless techniques, employing entanglement or coherence, generally leads to a decrease in the James-Stein advantage. In practice, noise precludes the achievement of theoretically optimal quantum improvements \cite{Demk2012}.  We show that in the presence of noise, the James-Stein advantage can be considerable. We demonstrate these results in the frequentist and Bayesian regimes. We also consider other experimental limitations, such as imperfect detectors. In such scenarios, it is advantageous to limit the number of measurements of quantum states via \textit{postselective filters} (see below).
These filters harness use quantum effects \cite{Arvidsson-Shukur2020} to compress the information from many copies of a parameterised quantum state, into (arbitrarily) fewer copies. By utilising such compression, one requires few measurements of quantum states to estimate an unknown parameter.  We construct an explicit iterative scheme for postselection of Gaussian states. Analysing this scheme, we find that the James-Stein advantage can be increased by postselective filtering. Moreover, postselective filtering can be improved by James-Stein estimation.

\section{Preliminaries}
\subsection{Frequentist parameter estimation}
Consider an unknown $n$-dimensional parameter  $\theta\in\Theta$, where $\Theta\subseteq\mathbb{R}^n$ is a known parameter space.  To estimate $\theta$, one can sample a random variable $X$ whose distribution $P_\theta$ has a probability density function (pdf) $f(x|\theta)$. If $X$ takes the value $x$, one denotes the corresponding estimate of $\theta$ by $\th(x)$. The function $\th$ is called an estimator. The performance of an estimator  is quantified by its risk:
\begin{equation}
	R(\th,\theta) = \mathbb{E}_{X\sim P_\theta}\left[\norm{\th(X)-\theta}^2\right].
\end{equation}
Using the risk, one can quantify the advantage of one estimator over another at a fixed value of the parameter $\theta$:
\begin{equation}
    \AD(\th[1], \th[2], \theta) = \frac{R(\th[2], \theta)}{R(\th[1], \theta)}.
\end{equation}
If $\AD(\th[1], \th[2], \theta')>1$, it is beneficial to use $\th[1]$ over $\th[2]$ when $\theta=\theta'$.\\

In the case of $N$ separate observations $X=(X_1,\dots, X_N)$, one considers a family of estimators $\th_N$; one for each value of $N$. The sequence $R(\th_N, \theta)$ quantifies how well an estimation strategy $\th_N$ performs with increasing resources. The most commonly studied regime in parameter estimation is the asymptotic limit, where $\theta$ is fixed and $N\to\infty$ \cite{Lehm1998, Brau1994, Alba2020}.  In rough terms, the ``best" possible estimator $\th[OPT]_N$ has risk scaling as 
\begin{equation}\label{eqn:classical_fisher_scaling}
    NR(\th[OPT]_N, \theta) \to \Tr[F(\theta)^{-1}],
\end{equation}
where $F(\theta)$ is the Fisher information matrix \cite{vaar1998}, given by
\begin{align}
    F(\theta)_{ij} &= \notag \\ &\mathbb{E}_{X\sim P_\theta}\left[\frac{\partial}{\partial\theta_i }\log f(X|\theta) \frac{\partial}{\partial\theta_j }\log f(X|\theta)\right].
\end{align}

In the quantum setting \cite{Hels1969}, a parameter $\theta$ is encoded in some quantum state $\rho(\theta)$. Often, one has $N$ copies of the same state $\sigma(\theta)$, in which case $\rho(\theta)=\sigma(\theta)^{\otimes N}$.
One can extract information by measuring $\rho(\theta)$. This involves choosing a POVM \cite{Neil2010}, $M = \{M_k\}$, where $M_k\geq 0$, and $\sum_k M_k = \mathbbm{1}$. The measurement and state induce a probability distribution parametrised by $\theta$:
\begin{equation}
	p_k^{(M)}(\theta) = \Tr[\rho(\theta)M_k].
\end{equation}
Once a measurement is chosen, the quantum-parameter-estimation problem has  been reduced to a classical problem (where $X$ takes values in the set of possible measurement outcomes). The quantum analogue of equation \eqref{eqn:classical_fisher_scaling} is given by 
\begin{equation}\label{eqn:quantum_fisher_scaling}
    NR(\th[qOPT]_N, \theta) \to C^H(\theta),
\end{equation}
where $C^H$ is called the Holevo-Cramer-Rao bound; see Ref. \cite{yang2019attaining} for details. Note that the optimal estimation strategy $\th[qOPT]_N$ includes optimisation over both measurements and estimators.

\subsection{The James-Stein Estimator}\label{sec:james_stein}
A particularly important parameter-estimation problem concerns data that is normally distributed around an unknown parameter: $Z\mid\theta\sim \mathcal{N}(\theta, \Sigma)$\footnote{Here, the notation $Z|\theta$ denotes a random variable $Z$ for a given value of $\theta$. Moreover, $\sim$ means ``distributed as".}. Here, $\mathcal{N}(\theta, \Sigma)$ denotes the (multivaritate) Gaussian (normal) distribution \cite{Vaar2000}, with mean $\theta\in\mathbb{R}^n$ and covariance matrix $\Sigma$. For simplicity we assume that $\Sigma$ is known, though this assumption can be dropped  \cite{Chet2012}. \\


For normally distributed data, $Z\mid\theta\sim \mathcal{N}(\theta, \Sigma)$, an intuitive estimator is the MLE, which aligns the estimate with the observed data: $\MLE(z) = z$. The MLE's risk is $R(\MLE, \theta) = \Tr(\Sigma)=\Tr[F(\theta)^{-1}]$. Below, we compare the MLE with two types of James-Stein estimators \cite{Li1988}.  The first is parameterised by a fixed vector $\nu\in\mathbb{R}^n$ and is given by
\begin{equation}\label{eqn:JS_estimator}
	\th[$\nu$-JS](z) = z -\frac{(n-2)\Sigma^{-1}}{(z-\nu)^T\Sigma^{-2}(z-\nu)}(z-\nu),
\end{equation}
where $n \geq 3$. 
$\th[$\nu$-JS](z)$ shrinks the Gaussian random variable $Z$ towards the vector $\nu$. The risk of $\th[$\nu$-JS](z)$ is
\begin{align}\label{eqn:JS_risk}
	R(\th[$\nu$-JS],\;&\theta) = \Tr(\Sigma) \notag\\
        &- (n-2)^2\mathbb{E}\left[\frac{1}{(Z-\nu)^T\Sigma^{-2}(Z-\nu)}\right],
\end{align}
where the expectation is taken over $Z\mid\theta\sim\mathcal{N}(\theta, \Sigma)$ \cite{Li1988}. The second term is strictly negative. Thus, $\AD(\JS, \MLE, \theta)>1$ for any value of $\theta$. The advantage of the James-Stein estimator is biggest when $\theta$ is close to $\nu$ and $\Sigma$ is ``large". Below, we fix $\nu = 0$, and let $\JS \equiv \th[0-JS]$. If one expects  $Z$ to be close to $\nu\neq 0$ (and hence that $\th[$\nu$-JS]$ will perform better than $\JS$), one can simply translate the origin to $\nu$. \\

Instead of shrinking towards a fixed vector, the second type of James-Stein estimator shrinks $Z$ towards its mean value. For $z\in\mathbb{R}^n$, let $z_m \equiv (1/n)(\sum_i z_i)(1,\dots, 1)$ be a vector whose entries are all the mean value of $z$. For $n\geq 4$, the second type of James-Stein estimator is then given by
\begin{equation}\label{eqn:cJS_estimator}
	\mJS(z) = z -\frac{(n-3)\Sigma^{-1}}{(z-z_m)^T\Sigma^{-2}(z-z_m)}(z-z_m),
\end{equation}
with risk
\begin{align}
	R(&\mJS,\theta) = \Tr(\Sigma) \notag\\
        &-(n-3)^2\mathbb{E}\left[\frac{1}{(Z-Z_m)^T\Sigma^{-2}(Z-Z_m)}\right].
\end{align}
Here, the expectation is evaluated over $Z\mid\theta\sim\mathcal{N}(\theta, \Sigma)$ \cite{Li1988}. The advantage of this James-Stein estimator is bigger the more isotropic $\theta$ is (and the ``larger" $\Sigma$ is). Even if there are no correlations between $Z$'s components, i.e., $\Sigma$ is diagonal,  both $\JS$ and $\mJS$ combine the components of $Z$ in a non-trivial manner. \\

Since the risks of $\JS$ and $\th[mJS]$ have similar forms, it is usually sufficient to consider $\JS$; $\th[mJS]$ will behave similarly. An exception is given by postselective metrology (section \ref{sec:postselection}), where we show that only $\th[mJS]$ is relevant. Otherwise, one should choose $\JS$ or $\th[mJS]$ depending on the expected properties of $\theta$: if $\theta$ is small,  use $\JS$; if $\theta$ is isotropic,  use $\th[mJS]$. If no information about $\theta$ is known, one can use either estimator, or a convex combination of them. \\
 
If one receives independent identically distributed normally distributed copies $Z_1,\dots Z_N$, then the sample mean $\bar{Z}_N$ is also normally distributed: $\bar{Z}_N\sim \mathcal{N}(\theta, \Sigma / N)$. Thus one can define $\MLE_N, \JS_N$ and $\mJS_N$: one simply applies each estimator to the sample mean. For the James-Stein estimators, this  involves modifying $\Sigma$ by a factor of $1/N$ in equations \eqref{eqn:JS_estimator} and \eqref{eqn:cJS_estimator}. One can see that
\begin{equation}\label{eqn:JS_advantage}
    \AD(\MLE_N, \JS_N, \theta) = 1 - \Theta(1/N),
\end{equation}
for both types of James-Stein estimator\footnote{We say that $f(x)=\Theta[g(x)]$ if $f$ and $g$ have the same asymptotic scaling, i.e. there exist constants $C, D\geq 0$ such that for $x$ sufficiently large $Cg(x)\leq f(x) \leq Dg(x)$.}. Thus, the James-Stein advantage diminishes with more data. The MLE and James-Stein estimators all saturate equation \eqref{eqn:classical_fisher_scaling}: they are asymptotically optimal.

\subsection{Gaussian states}
The most natural application of the James-Stein estimator to quantum metrology arises when measurement outcomes are normally distributed. Thus, we consider Gaussian states \cite{Weed2012}---states which are characterised by a ``mean" and ``variance", like a Gaussian distribution. More precisely, suppose that the Hilbert space $\mathcal{H}$ is a tensor product of $\ell$ single-mode Fock spaces. Examples include the Hilbert space of $\ell/3$ 3-dimensional particles, or a system with $q$ optical modes \cite{Weed2012}. There are $\ell$ pairs of canonical variables $Q_i,P_i,\; i=1,\dots,\ell$ satisfying the canonical commutation relations
\begin{equation}
	[Q_i,Q_j]=[P_i,P_j]=0,\quad [Q_i,P_j] = i\delta_{ij}.
\end{equation}
We group the canonical variables into a single vector $R=(Q_1,P_1,\dots,Q_\ell,P_\ell)$. For a state $\rho$ and vector $\xi\in\mathbb{R}^{2\ell}$ we define $\rho$'s characteristic function \cite{Weed2012} by
\begin{equation}
	\chi_\rho(\xi) = \Tr(\rho e^{i\xi^TR}).
\end{equation}
As in classical probability theory,  $\rho$ is completely characterised by $\chi_\rho(\xi)$ \cite{Englert2003}. A state $\rho$ is called Gaussian \cite{Demk2020} if there exists some vector $r\in\mathbb{R}^{2\ell}$ and some positive semi-definite matrix $A\geq0$ such that
\begin{equation}
	\chi_\rho(\xi) = e^{i\xi^Tr}e^{-\frac{1}{2}(\xi^TA\xi)}.
\end{equation}
This is exactly the characteristic function of a classical Gaussian probability distribution \cite{Vaar2000}, where $r$ is the mean of the distribution, and $A$ is its covariance matrix.\\

\section{Unknown Gaussian Channel}\label{sec:Gaussian_channel}
In this Section, we consider the application of the James-Stein estimator to estimating the mean of a Gaussian state, encoded by some quantum process. We assume that unknown parameters are encoded by a unitary $U(\theta)$ \cite{Giov2006}. In analogy with the number of classical samples from a probability distribution, we define $N$ as the number of applications of $U(\theta)$. Thus, $N$ quantifies experimental resources. \\

Let $\rho_0$ be a Gaussian state with mean zero and covariance matrix $A$, and let $U(\theta) = e^{i(\Omega \theta)^TR}$, where
\begin{equation}\label{eqn:swap_matrix_def}
    \Omega = \begin{pmatrix}
	\omega & &\\
	& \ddots &\\
	& &\omega
	\end{pmatrix} \text{, and } \omega=\begin{pmatrix}
	0 &1\\
	-1 &0
	\end{pmatrix}.
\end{equation}
 Then, we find that $\rho(\theta) = U(\theta)\rho_0 U(\theta)^\dag$ is a Guassian state with mean $\theta\in\mathbb{R}^{2\ell}$ and covariance matrix $A$.\\

 As $Q$ and $P$ are incompatible, we cannot simultaneously measure them. In Ref. \cite{Demk2020}, it was shown that this problem can be circumvented with the use of an ancilla Gaussian state $\tilde{\rho}$. The ancilla has the same Hilbert space as $\rho(\theta)$, with canonical variables $\tilde{Q}_i,\tilde{P}_i,\; i=1,\dots,\ell$. We prepare $\tilde{\rho}$ with mean 0 and some covariance matrix $\tilde{A}$. Then, we measure the state $\rho(\theta)\otimes\tilde{\rho}$ with respect to the commuting operators $ Q_i + \tilde{P}_i, P_i + \tilde{Q_i}, \; i=1,\dots,\ell$. \\

We define $\Sigma = A + \tilde{A}$; note that the Heisenberg uncertainty relation fundamentally lower bounds how small $\Sigma$ can be (see Ref. \cite{Demk2020} for details). In Appendix \ref{app:gaussian_mmnt}, we show, via a calculation using the Wigner function, that if $Z$ is a random variable corresponding to the outcome of the aforementioned measurement process, then $Z\mid\theta\sim\mathcal{N}(\theta, \Sigma)$. Since $Z$ has a Gaussian distribution we may use the James-Stein estimator. \\

If we are given $N$ copies of the unitary $U(\theta)$, a ``classical" estimation strategy is given by preparing $N$ copies of $\rho(\theta)$ and measuring each separately \cite{Giov2011}. One finds that the sample mean has distribution $\bar{Z}_N\sim \mathcal{N}(\theta, \Sigma / N)$ and thus that
\begin{equation}\label{eqn:default_mle_scaling}
    R(\MLE_N, \theta) = \Tr(\Sigma)/N.
\end{equation}
The James-Stein risk scales as in equation \eqref{eqn:JS_advantage}. \\

In Ref. \cite{Demk2020}, it was shown that upon optimisation of $\tilde{A}$, $\Tr(\Sigma) = C^H(\theta)$. Thus, the aforementioned measurement combined with the MLE saturates equation \eqref{eqn:quantum_fisher_scaling} and is asymptotically optimal. However, since $\AD(\JS, \MLE, \theta) > 1$ we note that $R(\JS_N,\theta) < C^H(\theta)/N$. Thus, for any finite $N$, the James-Stein estimator has risk \textit{smaller} than the asymptotic minimum, given in equation \eqref{eqn:quantum_fisher_scaling}. The James-Stein advantage must vanish as $N\to\infty$ [as required by equation \eqref{eqn:quantum_fisher_scaling}] but can be substantial for finite $N$; see Fig. \ref{fig:sequential_U_with_noise} (orange curve).\\

A common ``quantum" strategy \cite{smith2023adaptive, Smith22, Giov2006, Giov2011, Ji2008} is to apply $U(\theta)$ sequentially $N$ times to produce the state $\rho_N(\theta) = U^N(\theta)\rho_0 {U^N(\theta)^\dag}$. Now, $\rho_N(\theta)$ is a Gaussian state with covariance matrix $A$ and mean $N\theta$. Thus, letting $Z^{(N)}$ denote the outcome of measuring the mean of $\rho_N(\theta)$, we have $Z^{(N)}/N\mid\theta\sim\mathcal{N}(\theta,\Sigma/N^2)$. \\

We denote by $\th[qMLE]_N$ and $\th[qJS]_N$ the MLE and James-Stein estimators applied to $Z^{(N)}/N$, respectively. We find that
\begin{equation}\label{eqn:noiseless_advantage}
    \AD(\th[qMLE]_N, \MLE_N,\theta) = N.
\end{equation}
Using the unitary sequentially gives a factor of $N$ improvement over using it on separate copies of $\rho_0$. Similarly, $\AD(\th[qJS]_N, \JS_N,\theta) = \Theta(N)$, so that the James-Stein estimator is also improved. However, there is a corresponding decrease in the James-Stein advantage compared to equation \eqref{eqn:JS_advantage}:
\begin{equation}\label{eqn:noiseless_suppression}
    \AD(\th[qMLE]_N, \th[qJS]_N, \theta) = 1 - \Theta(1/N^2).
\end{equation}
In summary, whilst the James-Stein estimator still has an advantage over the MLE, this advantage is suppressed by coherent quantum effects, in the form of sequential unitary applications. See the dotted curve in Fig. \ref{fig:sequential_U_with_noise}. The use of an entangled probe state will show similar behaviours. 

\section{Gaussian Channels with Noise}\label{sec:noisy_Gaussian}

In Section \ref{sec:Gaussian_channel}, we assumed  perfect (noiseless) unitary evolution. In this Section, we lift this assumption and consider a common noise model. We show that the James-Stein and quantum-metrological advantages can be simultaneously large. \\

Suppose that there is coherent noise, such that we apply $U(\phi)$ instead of $U(\theta)$, where $\phi$ fluctuates randomly around $\theta$. This is modelled by the channel 
\begin{equation}
    \Lambda_\theta(\sigma) = \int d\phi f_\theta(\phi) U(\phi)\sigma U(\phi)^\dag,
\end{equation}
where $f_\theta$ is the pdf of a normal distribution with mean $\theta$ and covariance matrix $\Delta$. One can check that $\Lambda_\theta$  maps Gaussian states to Gaussian states: if $\rho$ is a Gaussian state with mean $r$ and covariance $A$, then $\Lambda_\theta(\rho)$ is a Gaussian state with mean $r+\theta$ and covariance $\Delta + A$. Channels that map Gaussian states to Gaussian states are also called Gaussian \cite{Eise2005}. \\

\begin{figure}
    \centering
    \includegraphics[width =\linewidth]{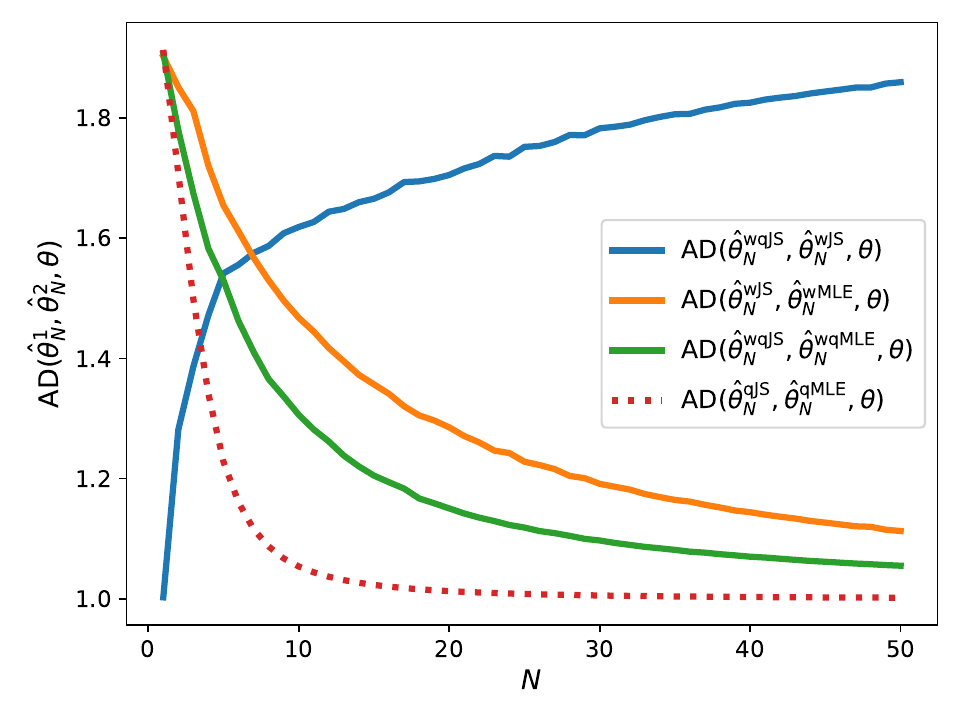}
    \caption{Comparison of various estimation strategies when $\theta = (0.5,-0.2,0.3,0.1)\in\mathbb{R}^4$, and $\Sigma, \Delta = 4\times \mathbbm{1}$. The James-Stein advantage with sequential unitary applications in the presence of noise (green curve) may be large. This contrasts to the case of sequential applications with no noise (dotted red line), where the James-Stein advantage quickly becomes negligible. The James-Stein advantage is even larger when applying $\Lambda_\theta$ on separate copies of $\rho(\theta)$ (orange curve), but in this case one loses the quantum advantage from sequential applications (demonstrated by the blue curve).}
    \label{fig:sequential_U_with_noise}
\end{figure}

We let $\bar{W}_N$ denote the sample mean of measuring $N$ copies of $\Lambda_\theta(\rho_0)$, and $W^{(N)}$ correspond to measuring a single copy of $\Lambda_\theta^N (\rho_0)$. We find that
\begin{align}
    \bar{W}_N\mid\theta &\sim \mathcal{N}(\theta, (\Sigma + \Delta)/N)\\
    W^{(N)}/N\mid\theta &\sim \mathcal{N}(\theta, \Sigma/N^2 + \Delta/N).
\end{align}
We label the estimation strategies $\MLE$, $\JS$, $\th[qMLE]$ and $ \th[qJS]$ applied to the (noisy) $W$ variables with a ``w" in superscript. We find that the unbounded asymptotic quantum-metrological advantage of equation \eqref{eqn:noiseless_advantage} is reduced to a constant factor:
\begin{equation}
    \AD(\th[wqMLE]_N , \th[wMLE]_N , \theta) = 1 + \frac{\Tr(\Sigma)}{\Tr(\Delta)} - \Theta(1/N). 
\end{equation}
Nevertheless, we stress that it can still be advantageous to sequentially apply $\Lambda_\theta$. Depending on the relative size of $\Sigma$ and $\Delta$, this advantage may be considerable. Moreover, we find that
\begin{equation}\label{eqn:noisy_suppression}
    \AD(\th[wqMLE]_N,\th[wqJS]_N, \theta) = 1 - \Theta(1/N).
\end{equation}
Comparing equations \eqref{eqn:noiseless_suppression} and \eqref{eqn:noisy_suppression}, we see that the James-Stein advantage is larger in the presence of noise. This is demonstrated in Fig. \ref{fig:sequential_U_with_noise} for a fixed $\theta\in\mathbb{R}^4$. The blue curve $\AD(\th[wqJS]_N, \th[wJS]_N, \theta)$ shows that sequentially applying $U(\theta)$ leads to a significant advantage, even in the presence of noise.  When $N=50$, we find that $\AD(\th[wqJS]_N, \th[wJS]_N, \theta) \approx 1.8$; a quantum metrological improvement of roughly $ 80\%$. From the other curves, we see that the James-Stein advantage falls off less rapidly when  $U(\theta)$ is applied sequentially with noise than without noise. When $N=25$, the James-Stein advantage is about 10\% when sequentially applying $U(\theta)$ with noise; it is 1\% when sequentially applying $U(\theta)$ without noise. We emphasise that these James-Stein advantages are ``free". No extra experimental resources are required, simply an alternative post-processing of the measured data.

\section{Bayesian Gaussians}\label{sec:Bayesian_Gaussian}
In Bayesian parameter estimation, one assumes that there is some prior distribution $\pi$ on the parameter space $\Theta$ that encodes an \textit{a priori} belief about the parameter to be estimated \cite{Wass2004}. With respect to this belief, one can  consider the Bayes risk of an estimator $\th$:
\begin{equation}\label{eqn:Bayes_risk}
    R_\pi(\th) = \mathbb{E}_{\vartheta\sim \pi}[R(\th, \vartheta)] = \int R(\th,\theta) d\pi(\theta).
\end{equation}
An estimator $\th$ is called Bayesian if it minimises the Bayes risk, as defined in equation \eqref{eqn:Bayes_risk}.\\

We consider the case of a normally distributed prior $\theta\sim\mathcal{N}(\theta_0,\Xi)$. To capture the different cases of Sections \ref{sec:Gaussian_channel} and \ref{sec:noisy_Gaussian}, we consider a random variable
\begin{equation}\label{eqn:bayesian_normal}
    Z_N\mid\theta\sim\mathcal{N}(\theta, \Gamma_N),
\end{equation}
where $\Gamma_N$ is some full-rank covariance matrix that depends on $N$. We denote the Bayesian, MLE and James-Stein estimators by $\th[bMLE]$ and $\th[bJS]$, respectively. In Appendix \ref{app:bayesian_estimator}, we show that the Bayes estimator for equation \eqref{eqn:bayesian_normal} is
\begin{equation}
    \tb(z) = (\Gamma_N^{-1} + \Xi^{-1})^{-1}(\Gamma_N^{-1} z + \Xi^{-1}\theta_0).
\end{equation}
We also calculate the Bayes risk of the MLE and James-Stein estimators (see appendix \ref{app:bayes_risks}); the results are summarised in Table \ref{tab:bayes_risks}.

\begin{table}[H]\centering
    \begin{tabular}{M{.2\linewidth} | M{.6\linewidth}}
        \th  & R_\pi(\th) \\ [5pt] \hline
        \th[bMLE] & \Tr(\Gamma_N) \\ [6pt]
        \th[bJS]  & \Tr(\Gamma_N) - (2q-2)^2 \mathbb{E}\left[\frac{1}{{Y_N}^T\Gamma_N^{-2}{Y_N}}\right]\\ [15pt]
        \tb  & \Tr(\Gamma_N) - \Tr\left[\Gamma_N^2(\Gamma_N+\Xi)^{-1}\right]\\ [6pt]
    \end{tabular}\\ \vspace{5pt}
    \caption{\label{tab:bayes_risks} Comparison of the Bayes risk of the maximum-likelihood, James-Stein and Bayes estimators. Here, $Y_N \sim \mathcal{N}(\theta_0, \Gamma_N + \Xi)$.}
\end{table}

By the definition of the Bayes estimator, and since the James-Stein estimator performs better everywhere than the MLE [see equation \eqref{eqn:JS_advantage}], we  have that
\begin{equation}\label{eqn:bayes_risk_comp}
    R_\pi(\tb) < R_\pi(\th[bJS]) < R_\pi(\th[bMLE]).
\end{equation}
Thus, the Bayes estimator $\tb$ is preferable. However, in practice, one may be unable to use $\tb$, since it depends explicitly on the covariance matrix $\Xi$, which may be unknown at the point of experiment. Additionally, there is no hope to estimate $\Xi$; $\theta$ is sampled according to $\pi$, and subsequently $\theta$ is encoded into $N$ copies of $U(\theta)$. Thus, the $N$ copies of $U(\theta)$ yield no information about $\Xi$. Consequently, the James-Stein estimator may be the best available alternative to the Bayes estimator. \\

If $\Gamma_N\to 0$ as $N\to\infty$, then the Bayesian advantage over the MLE is second order in $\Gamma_N$: $R_\pi(\tb) = R_\pi(\th[bMLE]) - O(\Gamma_N^2)$ . Thus, $R_\pi(\tb)/R_\pi(\th[bMLE]) = 1- O(\Gamma_N) \to 1$. Moreover, the faster, $\Gamma_N$ converges to 0, the faster $R_\pi(\tb)/R_\pi(\th[bMLE])$ will converge to 1. Therefore, sequential applications of $U(\theta)$ will move the MLE risk, and hence by equation \eqref{eqn:bayes_risk_comp}, the James-Stein risk, closer to the optimal Bayes estimator's risk. That is, the use of quantum-metrological techniques can move the James-Stein risk closer to the unattainable minimal Bayes risk.

\section{Postselected Guassian states} \label{sec:postselection}
 In this Section, we consider the application of postselection (defined below) \cite{Jenne2021, salvati2023compression, lupu2022negative, Arvidsson-Shukur2020} to the estimation of the mean of a Gaussian state. We also investigate how postselection affects the James-Stein advantage. For notational simplicity, we assume that the covariance matrix $\Sigma\propto\mathbbm{1}$ and that our state is pure. Additionally, by the ancilla trick described in Section \ref{sec:Gaussian_channel}, we may assume that $\theta$ is encoded solely in the positional degree of freedom of our state. That is, 
 \begin{equation}
    \ket{\psi_\theta} = C^{n/2}\int d^n x e^{-{\norm{x-\theta}^2/B}}\ket{x},
\end{equation}
for some known $n, B>0$, some unknown $\theta\in\mathbb{R}^n$, and $C = \sqrt{2/\pi B}$. For example, this is the state of $n\geq 4$ (continuous-variable) non-entangled, quantum sensors, each estimating some parameter $\theta_i\in\mathbb{R}$. Given $N$ copies of $\ket{\psi_\theta}$, as detailed in Section \ref{sec:Gaussian_channel}, one could measure the position of each state and take the sample mean to estimate $\theta$ with the corresponding risk scaling as $1/N$. \\

Many experiments suffer from limitations such as detector saturation \cite{luu2006saturation, wang2010saturation, frehlich1990effects, Glad2022} (if the intensity of photons arriving at a photon detector is too high, the detector performance degrades) and dead-times \cite{hachisu2015intermittent, wahl2020photon, rohde2006modelling} (a detector must reset for a brief time  between particle observations). Sometimes, one can only measure $M$ of the $N$ states $\ket{\psi_\theta}$ produced per unit time. It may be that $M\ll N$, so that the rate at which one can measure information is much less than the rate that one can produce it.\\

If one were to simply discard a random fraction $(1-M/N)$ of the copies of $\ket{\psi_\theta}$, the risk of our estimation strategies would scale as $1/M$---worse than $1/N$. However, if one knows that $\theta$ is a small displacement from some known value $\theta_0$, i.e. $\delta = \theta-\theta_0$ satisfies $\norm{\delta}^2 \ll M/N$, then the information carried by $N$ probe states can be losslessly compressed down to $M\ll N$ probe states \cite{Jenne2021}. This is similar in spirit to weak-value amplification techniques \cite{Hosten2008, Dixon2009, Turner2011, Pfeifer2011, Starling2010-1, Starling2010-2, Xiao-Ye2013, Magana2014, Strubi2013, Viza2013, Egan2012,Pusey2014, kunjwal2019anomalous}. \\

Instead of discarding states at random, one can utilise a more clever filter described by a 2-outcome POVM, with operators $F$ and $\mathbbm{1}-F$.
The outcome corresponding to $F$ (alt. $\mathbbm{1}-F$) has the probe pass (alt. not pass) the filter. Let $t\in[0,1]$ be a transmission parameter and take
\begin{equation}
    F = \mathbbm{1} - (1-t^2)\dyad{\psi_{\theta_0}}.
\end{equation}
 $F$ transmits all states perpendicular to $\ket{\psi_{\theta_0}}$, but only allows $\ket{\psi_{\theta_0}}$ through with probability $t^2$. As in \cite{Jenne2021}, we implement $F$ with the natural Kraus operator
\begin{equation}
    K = \mathbbm{1}-(1-t)\dyad{\psi_{\theta_0}}.
\end{equation}
If a state passes the filter, it will be in the state $K\ket{\psi_\theta}/\norm{K\ket{\psi_\theta}}$. In Appendix \ref{app:postselect_approx}, we show that if $t \gg \norm{\delta}$, then the post-filter state of the particle is 
\begin{equation}\label{eqn:approx_postselected_state}
    \frac{K\ket{\psi_\theta}}{\norm{K\ket{\psi_\theta}}}\approx C^{n/2}\int d^n x\; e^{-\norm{x-\theta_0-\delta/t}^2/B}\ket{x}
\end{equation}
Moreover, we show the probability of passing the filter is approximately $t^2$.\\

If we measure the probe's position after postselection, we  observe a random variable $X$, approximately normally distributed as
\begin{equation}
    X\mid\theta\sim\mathcal{N}(\theta_0 + \delta/t, [B/4]\mathbbm{1}).
\end{equation}
Due to the factor of $1/t$, $X$ is more sensitive in changes to $\theta$ than position measurements of $\ket{\psi_\theta}$. Thus $Y=t(X-\theta_0) + \theta_0$ is distributed as
\begin{equation}
    Y \mid \theta \sim \mathcal{N}(\theta, t^2[B/4]\mathbbm{1}).
\end{equation}
It takes (on average) $1/t^2$ copies of $\ket{\psi_\theta}$ until a state passes the filter, but the resulting state has a variance that is $t^2$ smaller than before. Thus, the information in $N$ copies of $\ket{\psi_\theta}$ is losslessly compressed into $M \approx t^2N$ copies of $K\ket{\psi_\theta}/\norm{K\ket{\psi_\theta}}$. Such lossless compression has been shown to require genuine quantum effects \cite{Arvidsson-Shukur2020, Jenne2021, Pusey2014}.\\

We proceed to investigate the advantage of the James-Stein estimator in experiments with post-selective filters. We model the aforementioned situation in which one is limited by a detector. If we can only measure $M$ states per unit time, we send $N=M/t^2$ copies of $\ket{\psi_\theta}$ through the filter. In this way, we ensure that (on average) $M$ states arrive at the detector per unit time. Thus, we can make $t$ arbitrarily small (in the approximation $M\ll N$).  \\

We start with with no prior knowledge about $\theta$ and hence no initial guess $\theta_0$. This means that  $\JS$ is a sub-optimal choice of James-Stein estimator, since we do not expect $\theta$ to be close to any \textit{a priori} known value. However, if the probes are measuring similar quantities, one could expect that the $\theta_i$'s would be similar and thus that $\th[mJS]$ would still perform well. Thus, we only consider $\th[mJS]$ in this section, and not $\JS$. Because we require that $t\gg \norm{\delta}$, one cannot decrease $t$ until one has a good estimate $\theta_0$ of $\theta$. We run an iterative strategy (see Appendix \ref{app:post_it_strategy} for the algorithm and Appendix \ref{app:rejection_sampling} for details on the numerics), where one estimates $\delta$ (using sample variance) and then decreases $t$ when one is confident that $\delta$ is sufficiently small. We consider using the MLE or James-Stein estimator to generate our estimates $\theta_0$ of $\theta$, and denote these strategies by $\th[pMLE]$ and $\th[pmJS]$, respectively. As a baseline, we compare this to measuring $N$ copies of $\ket{\psi_\theta}$ with no postselection ($t=1$), and using either the MLE $\MLE$ or James-Stein estimator $\mJS$.\\

\begin{figure}[t]
    \centering
    \begin{subfigure}{.49\textwidth}
        \centering
        \includegraphics[width=\textwidth]{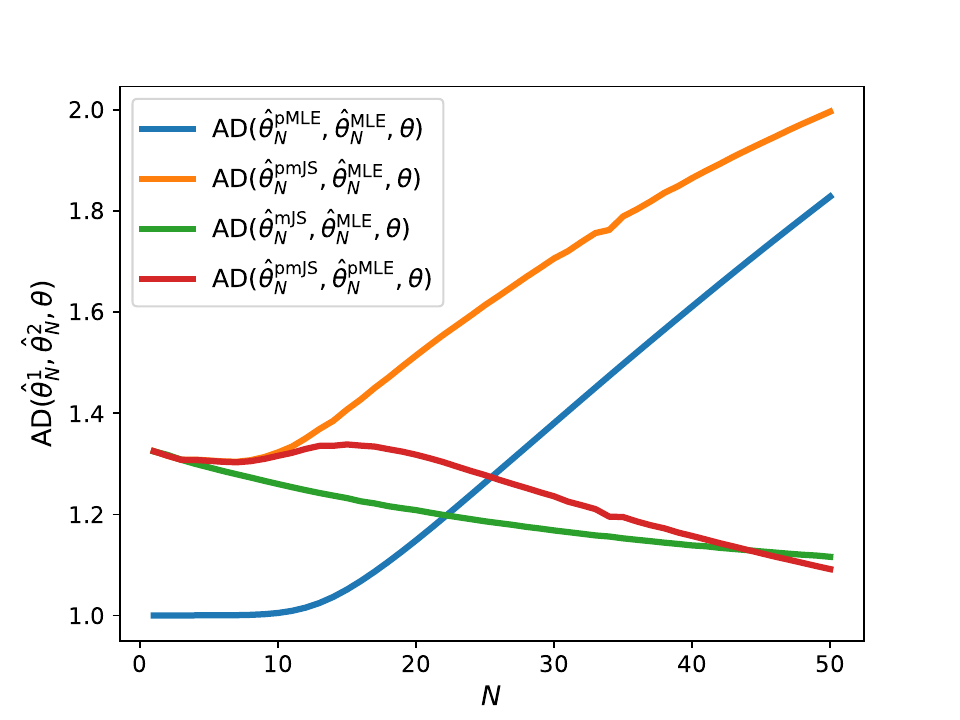}
        \caption{}
        \label{fig:postselection_fixed_theta}
    \end{subfigure}
    \hfill
    \begin{subfigure}{.49\textwidth}
        \centering
        \includegraphics[width=\textwidth]{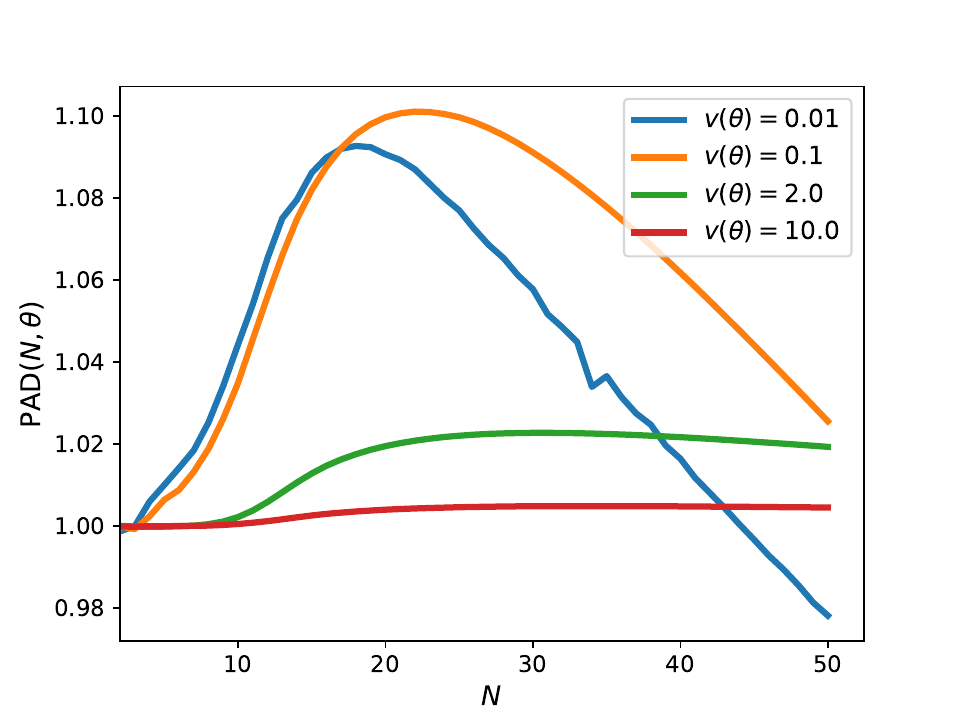}
        \caption{}
        \label{fig:postselection_varying_theta}
    \end{subfigure}
    \caption{(a) Comparison of various postselected and non-postselected estimation strategies when $\theta = (1, -2, 3, 1.5)/30$, $B=1$. (b) Comparison of the the affect of postselection on the James-Stein advantage for various $\theta$, $B=1$. The small kink in the blue line is due to computational shot noise.}
\end{figure}

In Fig. \ref{fig:postselection_fixed_theta}, we fix $\theta\in\mathbb{R}^4$. The blue curve shows that, when using the MLE, postselection gives an advantage. After some initial few measurements, this advantage increases linearly with the number of measurements. This is due to the increased sensitivity by $1/t$ in the postselected states, which leads to a more accurate estimate of $\theta$. The orange curve shows that the postselected James-Stein estimator also outperforms the non-postselected MLE. The green (alt. red) curve shows the James-Stein advantage without (alt. with) postselection. We see that using postselection the James-Stein advantage is initially increased, but eventually decays to below the non-postselected James-Stein advantage. This is because using $\th[mJS]$ allows one to postselect more quickly, so that the initial advantage increases. However, for larger values of $N$, $t$ shrinks, so the sample variance (and hence the James-Stein advantage) decreases. This behaviour is further analysed below. For $N < 40$, postselection boosts the James-Stein advantage.  \\

As discussed above, we expect $\th[mJS]$ to perform better when $\theta$ is more ``isotropic". Letting $\bar{\theta} = \sum \theta_i /n$, we quantify $\theta$'s isotropy  by 
\begin{equation}
    v(\theta) = \sum (\theta_i - \bar{\theta})^2
\end{equation}
To quantify the effect of postselection on the James-Stein advantage, we consider the ratio
\begin{equation}
    \PAD(N, \theta) = \frac{\AD(\th[pmJS]_N, \th[pMLE]_N, \theta)}{\AD(\th[mJS]_N, \th[MLE]_N, \theta)}.
\end{equation}
If $\PAD>1$, then postselection has increased the James-Stein advantage (compared to the non-postselected case). In Fig. \ref{fig:postselection_varying_theta}, we plot $\PAD(N, \theta)$ against $N$ for 4 different values of $\theta \in \mathbb{R}^4$, each with a different $v(\theta)$. As described above, the James-Stein estimator allows stronger postselection for smaller $N$, so that $\PAD >1$. However, as $N$ increases, the $1/t$ factor in sensitivity increases $v(\theta)$ by a factor of $1/t^2$, which degrades the James-Stein advantage. Thus, $\PAD$ eventually decreases. These effects are magnified when $v(\theta)$ is smaller.

\section{Conclusion}
We have introduced the James-Stein estimator to quantum metrology. We have focused on Gaussian shift estimation, but our techniques generalise to most common parameterised quantum states. We have shown that across a wide array of quantum-metrology protocols, the James-Stein estimator outperforms the MLE. This means that by simply changing post-processing techniques, one can increase experimental performance, without requiring any additional experimental resources. Our results highlight the non-trivial relationship between the asymptotic and non-asymptotic regimes of (quantum) metrology \cite{salmon2022classical, meyer2023quantum}; strategies that are asymptotically optimal may be sub-optimal with finite resources. We conclude that the James-Stein estimator can be a useful tool in quantum metrology. Moreover, quantum phenomena have a non-trivial relationship with the James-Stein advantage; they may increase it (sec. \ref{sec:postselection}), diminish it (sec. \ref{sec:Gaussian_channel}), or neither (sec. \ref{sec:noisy_Gaussian}). \\

\section*{Acknowledgements}
The authors wish to thank P. Karlsson for prompting them to investigate the James-Stein estimator in a quantum setting. Further, the authors thank N. Mertig, F. Venn, C. Long and J. Smith for their useful discussions and comments. W.S. was supported by the EPSRC and Hitachi. S.S. acknowledges support from the Royal Society University Research Fellowship. D.R.M.A.-S. was supported by Girton College.

\onecolumngrid

\twocolumngrid

\bibliography{bibliography}

\newpage
\appendix
\onecolumngrid

\section{Distribution of Gaussian measurement outcome}\label{app:gaussian_mmnt}
In this appendix, we find the distribution of the measurement described in Section \ref{sec:Gaussian_channel} (and Ref. \cite{Demk2020}). We begin by recapping the definition of a Gaussian state. Let $\mathcal{H}=\mathcal{F}^{\otimes \ell}$ be the tensor product of $\ell$ single mode Fock spaces. Let $R=(X_1,P_1,\dots,X_\ell,P_\ell)^T$ be a vector of operators, where $(Q_i,P_i)$ are the position and momentum canonical variables of the $i$th system, so that
\begin{equation}
[Q_i,P_j]=i\delta_{ij}.
\end{equation}

The characteristic function of a state $\rho$ is the function

\begin{equation}
	\chi_\rho:\mathbb{R}^{2\ell} \to \mathbb{C},\; \xi \mapsto \Tr(\rho e^{i\xi^TR}).
\end{equation}
A state is called Gaussian if for some $r\in\mathbb{R}^{2\ell}$ and $A$ some positive semi-definite matrix, we have that 
\begin{equation}
 \chi_\rho(\xi) = e^{i\xi^Tr}e^{-\frac{1}{2}(\xi^TA\xi)}.
\end{equation}
The Wigner function \cite{Tata1983} of the state is defined as the Fourier transformation of the characteristic function, that is 
\begin{equation}
	W_\rho:\mathbb{R}^{2\ell} \to \mathbb{C},\; x \mapsto \int \frac{d^{2\ell}\xi}{(2\pi)^{2\ell}} e^{-ix^T\xi}\chi_\rho(\xi). 
\end{equation}
If $\rho$ is a Gaussian state, then we find that
\begin{equation}
	W_\rho(x)\propto \exp(-\frac{1}{2}(x-r)^TA^{-1}(x-r)),
\end{equation}
i.e. a Gaussian in phase space. \\

As in Section \ref{sec:Gaussian_channel}, suppose we have a Gaussian state with a known covariance matrix $A$, but unknown mean $\theta$ that we wish to estimate. That is
\begin{equation}
	W_\rho(x) \propto \exp(-\frac{1}{2}(x-\theta)^TA^{-1}(x-\theta)).
\end{equation}

We cannot simultaneously measure the position and the momentum of the state, but  detailed in Ref. \cite{Demk2020}, we can circumvent this problem with the use of an ancilla. The ancilla is in a Gaussian state $\wt{\rho}$ which has the roles of position and momentum inverted. That is to say we define a vector of operators 
\begin{equation}
	\wt{R} = (\wt{P}_1,\wt{Q}_1,\dots,\wt{P}_\ell,\wt{Q}_\ell)^T,
\end{equation}
and ask that $\wt{\rho}$ satisfies
\begin{equation}
	\Tr(\wt{\rho} e^{i\xi^T\wt{R}}) = e^{i\xi^T\wt{r}}e^{-\frac{1}{2}\xi^T\wt{A}^{-1}\xi},
\end{equation}
for some vector $\wt{r}$ and covariance matrix $\wt{A}$.

Defining the swap matrix
\begin{equation}
	S = \begin{pmatrix}
	s & &\\
	& \ddots &\\
	& &s
	\end{pmatrix}, \quad
	s = \begin{pmatrix}
	0 &1\\
	1 &0
	\end{pmatrix},
\end{equation}
we find that $\wt{\rho}$ is a Gaussian state with mean vector $S \wt{r}$ and covariance matrix $S\wt{A}S$.

The joint state $\rho\otimes\wt{\rho}$ has Wigner function
\begin{equation}
	W_{\rho\otimes\wt{\rho}}(x) \propto \exp(-\frac{1}{2}(\bar{x}-\bar{\theta})^T\bar{A}^{-1}(\bar{x}-\bar{\theta})),
\end{equation}
where
\begin{equation}
	\bar{x} = \begin{pmatrix}
	\mathbbm{1} &\\ &S
	\end{pmatrix}x, \quad \bar{\theta} = \begin{pmatrix}
	\theta \\ \wt{r}
	\end{pmatrix},\quad \bar{A} = \begin{pmatrix}
	A &\\ &\wt{A}
	\end{pmatrix}.
\end{equation}

Ref. \cite{Demk2020} notes that the operators
\begin{equation}
	\{ Y_i = R_i + \wt{R}_i \}.
\end{equation}
mutually commute, and thus can be simultaneously measured. In fact, they show that this measurement scheme is optimal---the Fisher information of the resulting distribution is maximal. \\

Consider the probability density $f$ at the measurement outcome $Q_i + \wt{P}_i = \lambda_i$, $P_i + \wt{Q}_i = \mu_i$, denoted $f(\lambda_1,\mu_1,\dots,\lambda_\ell,\mu_\ell)$. Using standard properties of the Wigner function (see Ref. \cite{Tata1983} for details), we find
\begin{align}
	f(\lambda_1,\mu_1,\dots,\lambda_\ell,\mu_\ell) &= \int_{\mathbb{R}^{\ell}}d^{\ell}x\int_{\mathbb{R}^{\ell}}d^{\ell}\wt{x}\; W_{\rho\otimes\wt{\rho}}(x_1,\dots x_\ell,\wt{x}_1,\dots \wt{x}_\ell,\mu_1-\wt{x}_1,\dots,\mu_\ell-\wt{x}_\ell,\lambda_1-x_1,\dots,\lambda_\ell-x_\ell)\\
    &\propto \int_{\mathbb{R}^{\ell}}d^{\ell}x\int_{\mathbb{R}^{\ell}}d^{\ell}\wt{x} \; \exp(-\frac{1}{2}(\bar{x}-\bar{\theta})^T\bar{A}^{-1}(\bar{x}-\bar{\theta})),
\end{align}
where
\begin{equation}
	\bar{x} = (x_1,\mu_1-\wt{x}_1,\dots,x_\ell,\mu_\ell-\wt{x}_\ell,\lambda_1-x_1,\wt{x}_1,\dots,\lambda_\ell-x_\ell,\wt{x}_\ell)^T.
\end{equation}

We wish to extract the $\lambda$ and $\mu$ dependence from this integral. Let $x_0 = (x_1,-\wt{x}_1,\dots,x_\ell,-\wt{x}_\ell)^T$, $\mu_0 = (0,\mu_1,\dots,0,\mu_\ell)^T - \theta$ and $\lambda_0 = (\lambda_1,0,\dots,\lambda_\ell,0)^T-\wt{r}$. Using the definition of $\bar{A}$ and $\bar{\theta}$, we find that
\begin{equation}
	f(\lambda_1,\mu_1,\dots,\lambda_\ell,\mu_\ell) \propto \int_{\mathbb{R}^{\ell}}d^{\ell}x\int_{\mathbb{R}^{\ell}}d^{\ell}\wt{x}\; \exp(-\frac{1}{2}\left((x_0 +\mu_0)^TA^{-1}(x_0+\mu_0) + (\lambda_0-x_0)^T\wt{A}^{-1}(\lambda_0-x_0)\right)).\label{eqn:pdf}
\end{equation}

We complete the square in the exponent, i.e. write it in the form 
\begin{equation}
	-\frac{1}{2}\left((x_0 + r_0)^TM(x_0+r_0) + R\right),
\end{equation}
for some $r_0,R$. From inspection, we can see that
\begin{equation}
	M = A^{-1}+\wt{A}^{-1} \quad \text{and} \quad  r_0 = M^{-1}\left(A^{-1}\mu_0 - \wt{A}^{-1}\lambda_0\right).
\end{equation}

This gives
\begin{align}
	R &= \mu_0^TA^{-1}\mu_0 + \lambda_0^T\wt{A}^{-1}\lambda_0 - r_0^TMr_0,\label{eqn:unsimp_remainder}\\
	&= \mu_0^T M_1 \mu_0 + \lambda_0^T M_2 \lambda_0 + \lambda_0^T M_3 \mu_0 + \mu_0^T M_4 \lambda_0,\label{eqn:remainder}
\end{align}
where the matrices $M_i$ are given by
\begin{alignat}{2}
	M_1 &= A^{-1} - A^{-1}(A^{-1}+\wt{A}^{-1})^{-1}A^{-1}, \quad &&M_2 = \wt{A}^{-1} - \wt{A}^{-1}(A^{-1}+\wt{A}^{-1})^{-1}\wt{A}^{-1}, \nonumber\\
	M_3 &= \wt{A}^{-1}(A^{-1}+\wt{A}^{-1})^{-1}A^{-1},  &&M_4 = A^{-1}(A^{-1}+\wt{A}^{-1})^{-1}\wt{A}^{-1}.
\end{alignat}
In fact, all of the $M_i$ are equal, as we show in the following lemma.
\begin{lemma}\label{lem:matrix_simplification}
 $M_1=M_2=M_3=M_4= (A+\wt{A})^{-1}$.
\end{lemma}
\textbf{Proof: } By symmetry of the $M_i$ it suffices to prove the claim for $M_1$ and $M_3$. Note
\begin{equation}
	M_3 = (A(A^{-1}+\wt{A}^{-1})\wt{A})^{-1} = (A + \wt{A})^{-1}.
\end{equation}
Additionally,
\begin{align}
	M_1(A+\wt{A}) &= \mathbbm{1} + A^{-1}\wt{A} - A^{-1}((A^{-1}\wt{A}+\mathbbm{1})\wt{A}^{-1})^{-1}(\mathbbm{1}+A^{-1}\wt{A}),\\
	&= \mathbbm{1}.
\end{align}
The result follows. $\qed$

Combining equations (\ref{eqn:pdf}), (\ref{eqn:remainder}) and lemma \ref{lem:matrix_simplification}, we can make a change of coordinates in the integral to find
\begin{equation}
	f(z) = f(\lambda_1,\mu_1,\dots,\lambda_\ell,\mu_\ell) \propto \exp(-\frac{1}{2}\left((\lambda_0+\mu_0)^T(A+\wt{A})^{-1}(\lambda_0+\mu_0)\right)).
\end{equation}

Setting $\wt{r}$ to zero (as in \cite{Demk2020}), we reach
\begin{equation}
		Z|\theta \sim \mathcal{N}(\theta, A + \wt{A}),
\end{equation}
as claimed in Section \ref{sec:Gaussian_channel}.

\section{Bayesian Gaussian calculations}
In this appendix, we find the Bayes estimator for the measurement process described in Section \ref{sec:Bayesian_Gaussian}, as well as the Bayes risk of the maximum-likelihood, James-Stein and Bayes estimators

\subsection{Finding the Bayes estimator}\label{app:bayesian_estimator}
Recall that we are considering the (classical) Bayesian estimation problem of $Z_N \mid \theta \sim \mathcal{N}(\theta, \Gamma_N)$ and $\theta \sim \mathcal{N}(\theta_0, \Xi)$. Since we are using least squares loss \cite{Bane2005}, the Bayes estimator has the simple form
\begin{equation}
    \tb(z) = \mathbb{E}[\;\theta \mid z\;].
\end{equation}
Thus, consider the distribution of $\theta \mid z$.
\begin{equation}
    f(\theta | z) = \frac{f(z|\theta)f(\theta)}{\int f(z|\theta)f(\theta)d^{2n} \theta}.
\end{equation}
Note
\begin{equation}
    f(z|\theta)f(\theta) \propto e^{-\frac{1}{2}[(z-\theta)^T\Gamma_N^{-1}(z-\theta) + (\theta-\theta_0)^T\Xi^{-1}(\theta-\theta_0)]}\\
\end{equation}
We complete the square in the exponent, i.e. write it in the form
\begin{equation}
    (\theta-r_0)^T M (\theta-r_0) + R.
\end{equation}
From inspection, it is clear that we need
\begin{equation}
    M = \Gamma_N^{-1} + \Xi^{-1}, \qquad r_0 = M^{-1}(\Gamma_N^{-1} z + \Xi^{-1}\theta_0).
\end{equation}
We deduce 
\begin{equation}\label{eqn:conditional_theta_dist}
    \theta | z \sim \mathcal{N}(r_0 , M^{-1}),
\end{equation}
so that $\tb(z) = (\Gamma_N^{-1} + \Xi^{-1})^{-1}(\Gamma_N^{-1} z + \Xi^{-1}\theta_0)$.
We also find that
\begin{equation}
    R = z^T\Gamma_N^{-1} z + \theta_0^T\Xi^{-1}\theta_0 - r_0^T M r_0. 
\end{equation}
In analogy with equation \eqref{eqn:unsimp_remainder} we find that
\begin{equation}
    R = (z-\theta_0)(\Gamma_N + \Xi)^{-1}(z-\theta_0),
\end{equation}
And thus deduce
\begin{equation}\label{eqn:marginal_Z}
    Z_N\sim \mathcal{N}(\theta_0, \Gamma_N + \Xi).
\end{equation}

\subsection{Calculating Bayes risks}\label{app:bayes_risks}
We recall the three estimators we are comparing:
\begin{equation}
    \MLE(z) = z, \qquad \JS(z) = \left(\mathbbm{1}-(2q-2)\frac{\Sigma^{-1}}{z^T\Sigma^{-2}z}\right)z,\qquad \tb(z) = (\Gamma_N^{-1} + \Xi^{-1})^{-1}(\Gamma_N^{-1} z + \Xi^{-1}\theta_0).
\end{equation}
Since $R(\MLE, \theta) = \Tr(\Gamma_N)$, it is easy to see $R_\pi(\MLE) = \Tr(\Gamma_N)$.\\

From equation \eqref{eqn:JS_risk} we know 
\begin{equation}
    R(\JS,\theta) = \Tr(\Gamma_N) -  (2q-2)^2\mathbb{E}\left[\frac{1}{Z_N^T\Gamma_N^{-2}Z_N}\middle|\; \theta\; \right],
\end{equation}
By definition,
\begin{equation}
    R_\pi(\JS) = \mathbb{E}_{\theta\sim\pi}[R(\JS,\theta)],
\end{equation}
so using the tower law of expectation we find
\begin{equation}
    R_\pi(\JS) = \Tr(\Gamma_N) -  (2q-2)^2\mathbb{E}\left[\frac{1}{Z_N^T\Gamma_N^{-2}Z_N} \right],
\end{equation}
where the expectation is now taken over the full distribution of $Z_N$ that we found in equation \eqref{eqn:marginal_Z}.

It remains to find the Bayes-risk of the Bayes estimator $\tb$. Using by the tower-law of expectation,
\begin{align}
    R_\pi(\tb) = \mathbb{E}_{Z_N}\left(\mathbb{E}_{\theta|Z_N}\left[\norm{\theta-\tb(Z_N)}^2\middle|\;Z_N\;\right]\right)
\end{align}
Since $\tb(z)=\mathbb{E}[\;\theta\mid z\;]$ and $\theta \mid z$ is distributed according to equation \eqref{eqn:conditional_theta_dist}, the inner expectation is the expected squared distance of a multivariate Gaussian distribution from its mean, which is the trace of the covariance matrix: $\Tr(\Gamma_N^{-1} + \Xi^{-1})^{-1}$. The outer expectation is then the expectation of a constant so that
$R_\pi(\tb) = \Tr(\Gamma_N^{-1} + \Xi^{-1})^{-1} $.\\

In order to compare the Bayes risk of the Bayes estimator with the MLE and JS risks, it is useful to rewrite this risk using lemma \ref{lem:matrix_simplification}. Taking $A=\Gamma_N^{-1}$ and $\wt{A} = \Xi^{-1}$. We find
\begin{equation}
    R_\pi(\tb) = \Tr(\Gamma_N) - \Tr(\Gamma_N(\Gamma_N + \Xi)^{-1}\Gamma_N).
\end{equation}
Noting that $\Gamma_N$ is symmetric, and $\Gamma_N + \Xi$ is positive definite, we deduce that that $R_\pi(\tb)< R_\pi(\MLE)$.

\section{Postselection}
\subsection{Approximate postselected state}\label{app:postselect_approx}
In this appendix, we justify equation \eqref{eqn:approx_postselected_state} in Section \ref{sec:postselection}. First we calculate the overlap 
\begin{align}
    \braket{\psi_{\theta_0}}{\psi_\theta} &= C^n\int_{-\infty}^{\infty} d^n x\;\; e^{-(\norm{x-\theta}^2 + \norm{x-\theta_0}^2)/B},\\
    &= C^n\int_{-\infty}^{\infty} d^n x\;\; e^{-(2\norm{x-\theta/2 - \theta_0/2}^2 + \norm{\theta-\theta_0}^2/2)/B},\\
    &= e^{-\norm{\delta}^2/2B}.
\end{align}
Then
\begin{align}
    K\ket{\psi_\theta} &= \ket{\psi_\theta} - (1-t)e^{-\norm{\delta}^2/2B}\ket{\psi_{\theta_0}},\\
                    &= C^{n/2}\int d^nx\; (e^{-\norm{x-\theta}^2/B}-(1-t)e^{-\norm{\delta}^2/2B}e^{-\norm{x-\theta_0}^2/B})\ket{x},\\
                    &= C^{n/2}\int d^ny\; (e^{-\norm{y-\delta}^2/B}-(1-t)e^{-\norm{\delta}^2/2B}e^{-\norm{y}^2/B})\ket{y+\theta_0}, \label{eqn:post_filter_state}\\
                    &= C^{n/2}\int d^ny\; e^{-\norm{\delta}^2/B}e^{-\norm{y}^2/B}[e^{2y^T\delta/B}-(1-t)e^{\norm{\delta}^2/2B}]\ket{y+\theta_0}.
\end{align}
We can see from equation \eqref{eqn:post_filter_state} that the integrand only has effective support for $y=O(\delta)\ll 1$. Thus we can Taylor expand in $\delta$, treating $y$ as $O(1)$.
\begin{align}
    K\ket{\psi_\theta} &= C^{n/2}\int d^3y\; e^{-\norm{\delta}^2/B}e^{-\norm{y}^2/B}[t + 2y^T\delta/B + O(\norm{\delta}^2) ]\ket{y+\theta_0},\\
    &= tC^{n/2}\int d^ny\; e^{-\norm{\delta}^2/t^2B}e^{-\norm{y}^2/B}e^{2y^T\delta/tB}[1+O( \norm{\delta}^2/t^2)]\ket{y+\theta_0},\\
    &= tC^{n/2}\int d^ny\; e^{-\norm{y-\delta/t}^2/B}[1+O( \norm{\delta}^2/t^2)]\ket{y+\theta_0},\\
    &= tC^{n/2}\int d^nx\; e^{-\norm{x-\theta_0-(\theta-\theta_0)/t}^2/B}[1+O( \norm{\delta}^2/t^2)]\ket{x}.
\end{align}
So for $\norm{\delta}\ll t$, we find, as claimed in the main text, that
\begin{equation}
    \mathbb{P}(\text{pass filter}) = \norm{K\ket{\psi_\theta}}^2\approx t^2, \qquad \frac{K\ket{\psi_\theta}}{\norm{K\ket{\psi_\theta}}}\approx C^{n/2}\int d^nx\; e^{-\norm{x-\theta_0-(\theta-\theta_0)/t}^2/B}\ket{x}.
\end{equation}

\subsection{Iterative Strategy}\label{app:post_it_strategy}
In this appendix, we describe our iterative strategy for estimating $\theta$ using postselection. We wish to compare the James-Stein estimator and MLE. In order to capture both cases, we describe the algorithm in more general terms, referring only to an estimator $\th$ for estimating the mean of a Gaussian distribution. If one wishes to implement a specific strategy, one can replace $\th$ by, for example, $\MLE$ or $\mJS$.\\

Initially, we set $\theta_0 = 0$ and $t=1$. \\

Suppose we are on the $k$th measurement; $\theta_0$ and $t$ have been set to some known values. Recall (from appendix \ref{app:postselect_approx}), that the postselected state is roughly

\begin{equation}
    \frac{K\ket{\psi_\theta}}{\norm{K\ket{\psi_\theta}}}\approx C^{n/2}\int d^nx\; e^{-\norm{x-\theta_0-(\theta-\theta_0)/t}^2/B}\ket{x}.
\end{equation}

We measure the position of the postselected state, let the outcome be $X_k$. Rescaling, we take $Y_k = \theta_0 + t(X-\theta_0)$ so that $Y\sim \mathcal{N}(\theta, t^2(B/4)\mathbbm{1})$. We make an estimate $\th_k(Y)$ of $\theta$. Our collated estimate (using all the data so far) is given by $\bar{\theta}_k$, the sample mean of $\th_1,\dots,\th_k$. Denote the sample standard deviation of $\th_1,\dots \th_k$ by $\hat{\sigma}_k$. We estimate $\delta$ by $\hat{\delta}_k = \hat{\sigma}_k/\sqrt{k}$. \\

If $\delta_k$ is ever less than $0.3t$, we set $t$ to $3\hat{\delta}_k$ and $\theta_0$ to $\bar{\theta}_k$: our current best estimate of $\theta$. \\

If the strategy terminates after measurement $K$, we output $\bar{\theta}_K$. Thus, after measurement $k$, the current error of the estimation strategy is given by $R_k = \norm{\bar{\theta}_k -\theta}^2$. To estimate the risk of a particular strategy $\th=\mJS$ or $\th=\MLE$ at measurement $k$, we average $R_k$ over many individual repetitions.

\subsection{Exact sampling from the postselected pdf}\label{app:rejection_sampling}
In order to numerically simulate the postselection, we must sample exactly from the distribution 

\begin{equation}
    f_{\theta_0, t}(x) = \frac{|\matrixel{x}{K}{\psi_\theta}|^2}{\norm{K\ket{\psi_\theta}}^2},
\end{equation}
given by measuring the position of the postselected state. \\

In order to do this, we use rejection sampling. We give a brief summary of rejection sampling here, for details see Ref. \cite{casella2004generalized}. Assume there is a target pdf $f(x)$ that we wish to sample from. Rejection sampling requires a pdf $g(x)$ that one can sample from, and constant $M$ satisfying $f(x)\leq Mg(x)$. The rejection sampling algorithm is:
\begin{enumerate}
    \item Generate a candidate $x$ from the pdf $g(x)$.
    \item Output this candidate with probability $f(x)/Mg(x)$. If the candidate is rejected, return to step 1.
\end{enumerate}
Note that the chance of acceptance is given by $\int g(x) [f(x)/ Mg(x)] dx = 1/M$. Thus, on average, $M$ samples from $g(x)$ are required to generate one sample from $f(x)$.   \\

To use rejection sampling, we must find a constant $M_{\theta_0, t}$ and a pdf $g_{\theta_0, t}(x)$ that is easy to sample from, such that $f_{\theta_0, t}(x) \leq M_{\theta_0, t}g_{\theta_0, t}(x)$. First, we calculate 
\begin{align}
    \innerproduct{\psi_\theta}{\psi_{\theta_0}} = e^{-\norm{\delta}^2/2B},
\end{align}
where $\delta = \theta - \theta_0$. Then, recalling $K = (t-1)\dyad{\theta_{\theta_0}} + \mathbbm{1}$, we find
\begin{align}\label{eqn:f_numerator}
    |\matrixel{x}{K}{\psi_\theta}|^2 = C^n|e^{-\norm{x-\theta}^2/B}-(1-t)e^{-\norm{\delta}^2/2B}e^{-\norm{x-\theta_0}^2/B}|^2.
\end{align}
By expanding the square and integrating each of the resulting Gaussians, we find that
\begin{align}
    \norm{K\ket{\psi_\theta}}^2 &= \int dx |\matrixel{x}{K}{\psi_\theta}|^2, \\
    &= 1 + (t^2-1)e^{-\norm{\delta}^2/B}.
\end{align}
If $A,B\geq 0$, note that $|A-B|^2\leq A^2 + B^2$. Thus, we may bound equation \eqref{eqn:f_numerator} by
\begin{equation}
    |\matrixel{x}{K}{\psi_\theta}|^2 \leq C^n(e^{-2\norm{x-\theta}^2/B} + (1-t)^2e^{-\norm{\delta}^2/B}e^{-2\norm{x-\theta_0}^2/B})
\end{equation}

Thus,
\begin{align}
    f_{\theta_0, t}(x) &\leq C^d\frac{e^{-2\norm{x-\theta}^2/B} + (1-t)^2e^{-\norm{\delta}^2/B}e^{-2\norm{x-\theta_0}^2/B}}{1 + (t^2-1)e^{-\norm{\delta}^2/B}},\\
    &= \frac{1 + (1-t)^2e^{-\norm{\delta}^2/B}}{1 + (t^2-1)e^{-\norm{\delta}^2/B}}\left(\frac{1}{1 + (1-t)^2e^{-\norm{\delta}^2/B}}C^d e^{-2\norm{x-\theta}^2/B} + \frac{(1-t)^2e^{-\norm{\delta}^2/B}}{1 + (1-t)^2e^{-\norm{\delta}^2/B}}C^n e^{-2\norm{x-\theta_0}^2/B}\right)\\
    &= M_{\theta_0, t}g_{\theta_0, t}(x),
\end{align}
where
\begin{equation}
    M_{\theta_0, t} = \frac{1 + (1-t)^2e^{-\norm{\delta}^2/B}}{1 + (t^2-1)e^{-\norm{\delta}^2/B}},
\end{equation}
and
\begin{equation}
    g_{\theta_0, t}(x) = \frac{1}{1 + (1-t)^2e^{-\norm{\delta}^2/B}}C^n e^{-2\norm{x-\theta}^2/B} + \frac{(1-t)^2e^{-\norm{\delta}^2/B}}{1 + (1-t)^2e^{-\norm{\delta}^2/B}}C^n e^{-2\norm{x-\theta_0}^2/B}.
\end{equation}
Note that $g_{\theta_0, t}(x)$ is a pdf given by the convex sum of the pdf of two different normal distributions. Specifically, let 
\begin{equation}
    \phi_{\mu}(x) = C^n e^{-2\norm{x-\mu}^2/B},
\end{equation}
denote the pdf of a normal distribution with mean $\mu$ and covariance matrix $\frac{B}{4}\mathbbm{1}$. Then,
\begin{equation}
    g_{\theta_0, t}(x) = p_{\theta_0, t}\phi_{\theta}(x) + (1-p_{\theta_0, t})\phi_{\theta_0}(x),
\end{equation}
where
\begin{equation}
    p_{\theta_0, t} = \frac{1}{1 + (1-t)^2e^{-\norm{\delta}^2/B}}.
\end{equation}
Thus it is easy to sample from $g_{\theta_0, t}(x)$: sample from $\phi_{\theta}(x)$ with probability $p_{\theta_0, t}$, otherwise sample from $\phi_{\theta_0}(x)$. Therefore, we can use rejection sampling to sample from $f_{\theta_0, t}$. Note that as $t,\delta\to0, M\to\infty$, so as postselection becomes stronger, rejection sampling takes longer (on average) to run. 
\end{document}